\documentclass[preprint,eqsecnum,showpacs,tightenlines,amsmath,%
amssymb,nofootinbib,showkeys,byrevtex]{revtex4}
\usepackage{bm}

\begin{document}


\title{Spectral analysis of a flat plasma sheet model}

\author{M.~Bordag}
\affiliation{Institute for Theoretical Physics, University of
Leipzig, Augustusplatz 10/11, 04109 Leipzig, Germany}
\email{Michael.Bordag@itp.uni-leipzig.de}

\author{I.~G.~Pirozhenko}
\affiliation{Bogoliubov Laboratory of Theoretical Physics, Joint
Institute for Nuclear Research, 141980 Dubna, Russia}
\email{pirozhen@thsun1.jinr.ru} \email{nestr@thsun1.jinr.ru}
\author{V.~V.~Nesterenko}
\affiliation{Bogoliubov Laboratory of Theoretical Physics, Joint
Institute for Nuclear Research, 141980 Dubna, Russia}
\email{nestr@thsun1.jinr.ru}

\date{\today}


\begin{abstract}
The spectral analysis of the electromagnetic field on the background of a infinitely thin
flat plasma layer is carried out. This model is loosely imitating a single base plane
from graphite and it is of interest for theoretical studies of fullerenes. By making use
of the Hertz potentials the solutions to Maxwell equations with the appropriate matching
conditions at the plasma layer are derived and on this basis the spectrum of
electromagnetic oscillations is determined. The model is naturally split into the
TE-sector and TM-sector. Both the sectors have positive continuous spectra, but the
TM-modes have in addition a bound state, namely, the surface plasmon. This analysis
relies on the consideration of the scattering problem in the TE- and TM-sectors. The
spectral zeta function and integrated heat kernel are constructed for different branches
of the spectrum in an explicit form. As a preliminary, the rigorous procedure of
integration over the continuous spectra is formulated by introducing the spectral density
in terms of the scattering phase shifts.  The asymptotic expansion of the integrated heat
kernel at small values of the evolution parameter is derived. By making use of the
technique of integral equations, developed earlier by the same authors, the local heat
kernel (Green's function or fundamental solution) is constructed also. As a by-product, a
new method is demonstrated  for deriving the fundamental solution to the heat conduction
equation (or to the Schr\"odinger equation) on an infinite line with the $\delta $-like
source. In particular, for the heat conduction equation on an infinite line with the
$\delta$-source a nontrivial counterpart is found, namely, a spectral problem with point
interaction, that possesses the same integrated heat kernel while the local heat kernels
(fundamental solutions) in these spectral problems are different.
\end{abstract}



\maketitle
\section{Introduction}
\label{sec1} Now  the  employment  of  the fullerenes in  practical  activity becomes  of
rising importance (in superconducting technique,  in mechanical  engineering for reducing
the friction,  in  medicine for drug production and so on).
Fullerenes~\cite{Buckminster,Dresselhaus} are the third crystal modification  of carbon.
First they were produced  in  1980-ties (Nobel price in Chemistry in 1996,  R.~F.~Curl,
H.~W.~Kroto, and R.~E.~Smalley).  The elementary blocks of the fullerenes are the giant
carbon molecules (C$_{60}$, C$_{70}$, C$_{84}$, and greater) having the form of empty
balls, ellipsoids, tubes and so on. There appears a special branch of technology,
nanotechnology~\cite{Kadish}, which  is based on using the fullerenes and especially,
nanotubes. In theoretical studies of the fullerenes it is important to estimate the
ground state energy (the Casimir energy)  of such molecules. The carbon shells, forming
the structure of fullerenes, can be considered as infinitely thin layers, and their
collective electrons can be treated  as two-dimensional plasma layer by making use of the
hydrodynamic picture for its description. As  a result, the problem reduces  to the
solution  of the  Maxwell equations with charges  and currents distributed  along the
surfaces modeling the carbon sheets.  The method of boundary (matching) conditions
applies here, i.e., the Maxwell equations are considered only outside the plasma  layers
and  when approaching  these sheets the electric  and magnetic fields  should  meet
appropriate matching conditions. From  the mathematical stand point the pertinent
spectral problem proves to  be  well posed.

For constructing the quantum version of this model and, first of all, for a rigorous
treatment of the divergencies encountered here the spectral analysis of the underlying
field theory should be accomplished, i.e., the spectrum of the dynamical system in
question should be determined, the mathematically consistent  integration procedure over
this spectrum should be defined and on this basis the spectral zeta function and heat
kernels (integrated and local ones) should be calculated as well as their asymptotic
expansions under small values of the evolution parameter should be found.


The present work is exactly  devoted to the spectral analysis of an infinitely thin
plasma sheet model with the simplest geometry, namely, the plasma layer has the form of a
plane. The spectrum of the model proves to be very reach, namely, it contains  both
continuous branches and bound states (surface plasmon), and we managed to construct all
the pertinent spectral functions possessing interesting properties.

 The outline of the paper is as  follows. In Sec.~\ref{sec2} the physical
formulation of the model under study is briefly given. In Sec.\ \ref{sec3} the solutions
to the Maxwell equations obeying the pertinent matching conditions at the plasma layer
are constructed. The use of the Hertz vectors for this aim proves to be very effective.
Proceeding from this the spectrum of electromagnetic oscillations in the model is found.
As usual in electrodynamical problems the model is naturally separated into the TE-sector
and TM-sector. Both the sectors have positive continuous spectra, but the TM-modes have
in addition a bound state, namely, the surface plasmon. This analysis relies on the
consideration of the scattering problem in the TE- and TM-sectors. In Sec.~\ref{sec4} the
spectral zeta function and integrated heat kernel are constructed for different branches
of the spectrum in an explicit form. As a preliminary, the rigorous procedure of
integration over the continuous spectra is formulated by introducing the spectral density
(the function of the spectral shift) in terms of the scattering phase shifts.
Sec.~\ref{sec5}  is devoted to the construction of the local heat kernel (Green's
function or fundamental solution) in the model at hand. Here the technique of integral
equations governing the heat kernel to be found is used. As a by-product,  a new method
is demonstrated here for deriving the fundamental solution to the heat conduction
equation (or to the Schr\"odinger equation)  on an infinite line with the $\delta $-like
source. In particular, for the heat conduction equation on an infinite line with the
$\delta$-source a nontrivial counterpart is found, namely, a spectral problem with point
interaction, that possesses the same integrated heat kernel. However the local heat
kernels in these spectral problems are different. In the Conclusion (Sec.\ \ref{sec6})
the obtained results are summarized briefly.

\section{Formulation of the model}
\label{sec2} In recent papers~\cite{Barton1,Barton2} Barton has proposed  and
investigated the model of infinitesimally thin two-dimensional plasma layer that is
loosely inspired by considering a single base plane from graphite or the giant carbon
molecule C$_{60}$. Effectively the model is described by the Maxwell equations with
charges and currents distributed along the surface $\Sigma $ :
\begin{equation}
\label{eq-1} \bm{\nabla} \cdot \mathbf{B} =0, \quad \bm{\nabla}
\times \mathbf{E} -i\omega \mathbf{B}/c=0\,{,}
\end{equation}
\begin{equation}
\label{eq-2} \bm{\nabla} \cdot \mathbf{E} =4\pi \delta
(\mathbf{x}-\mathbf{x}_{\Sigma})\,\sigma, \quad  \bm{\nabla} \times
\mathbf{B} +i\omega
\mathbf{E}/c=4\pi\delta(\mathbf{x}-\mathbf{x}_{\Sigma})\,\mathbf{J}/c\,{.}
\end{equation}
It is assumed that the time variation of all the dynamical variables
is described by a common factor $e^{-i\omega t}$.

The properties of the plasma layer (for example, plasma  oscillations and screening)
differ considerably from those in bulk media~\cite{Fetter}. In hydrodynamic approach the
plasma is considered as the electron fluid embedded in a rigid uniform positive
background. Electrical neutrality requires that the equilibrium electron charge density
$-e n_0$ precisely cancels that of the background. In a dynamical situation, the electron
density is altered to $n(\mathbf{x},t)$, and the equation of continuity requires
\begin{equation}
\label{eq-3} \frac{\partial n}{\partial
t}+\bm{\nabla}\cdot(n\,\dot{\bm{\xi}} )=0\,{,}
\end{equation}
where $\dot{\bm{\xi}}(\mathbf{x},t)$ is the electron velocity at the
point $\mathbf{x}$. Usually one considers  the {\it linear} response
of an initially stationary system to an applied perturbation. In this
case, the induced charge density
\begin{equation}
\label{eq-4} \sigma(\mathbf{x},t)\equiv
-e\,(n(\mathbf{x},t)-n_0)\,{,}
\end{equation}
the velocity $\dot{\bm{\xi}}$, and the fields $\mathbf{E}$ and
$\mathbf{B}$ are of first order. In this approximation the continuity
equation (\ref{eq-3}) gives for the plasma layer
\begin{equation}
\label{eq-5} \dot{\sigma}
-en_0{\bm{\nabla}}_{\|}\cdot\dot{\bm{\xi}}=0\quad \text{or} \quad
\sigma = en_0{\bm{\nabla}}_{\|}\cdot\bm{\xi}\,{.}
\end{equation}
The superscripts $\|$ and $\bot$  indicate here and below the vector
components respectively parallel and normal to the surface $\Sigma$.

For the induced charge current $\mathbf{J}$ we have
\begin{equation}
\label{eq-6} \mathbf{J}=-en_0\dot{\bm{\xi}}=i\,e\,\omega \,n_0
\bm{\xi}\,{.}
\end{equation}
The Newton second law applied to an individual electron gives
\begin{equation}
\label{eq-7}
m\,\ddot{\bm{\xi}}(\mathbf{x},t)=-e\,\mathbf{E}_{\|}(\mathbf{x},t),\quad
\mathbf{x}\in \Sigma,\quad \text{or}\quad
\bm{\xi}=\frac{e}{m\omega^2}\,\mathbf{E}_{\|}\,{.}
\end{equation}
Finally, the induced-on-the-surface charges and electric currents
are determined by the parallel components of electric filed
\begin{equation}
\label{eq-7-0} \sigma=\frac{e^2n_0}{m\omega^2}\, \bm{\nabla}_{\|}\cdot
\mathbf{E}_{\|},\quad \mathbf{J}=i\,\frac{e^2n_0}{m\omega}\,\mathbf{E}_{\|}\,{.}
\end{equation}

As known~\cite{Stratton}, electromagnetic field, generated by charges and currents with
singular densities is described mathematically in the following way: outside the
singularities the Maxwell equations (\ref{eq-1}), (\ref{eq-2}) without sources should be
satisfied
\begin{eqnarray}
\label{eq-7a} \bm{\nabla} \cdot \mathbf{B} =0,&\quad& \bm{\nabla} \times \mathbf{E}
-i\omega \mathbf{B}/c=0\,{,} \\\label{eq-7b} \bm{\nabla} \cdot \mathbf{E} =0, &\quad&
\bm{\nabla} \times \mathbf{B} +i\omega \mathbf{E}/c=0,\quad \mathbf{x}\not\in \Sigma\,{,}
\end{eqnarray}
and  when approaching at singularities the limiting values of the fields  should meet the
matching conditions determined by the singular sources. In the regions of singularities the
Maxwell equations are not considered. The singular charges and currents in right-hand sides
of the Maxwell equations (\ref{eq-2}) lead to the following matching conditions
\begin{eqnarray}
[ \mathbf{E}_{\|}]=0, \quad [\mathbf{E}_{\bot}]&=&2 q(c/\omega)^2
 {\bm{\nabla}}_{\|}\cdot \mathbf{E}_{\|}\,{,} \label{eq-8} \\
{} [\mathbf{B}_{\bot}] = 0, \quad [\mathbf{B}_{\|}]&=&-2i q(c/\omega)\, \mathbf{n}\times
\mathbf{E}_{\|}\,{.} \label{eq-9}
\end{eqnarray}
Here $q$ is a characteristic wave number $q=2\pi n e^2/mc^2$,  the square brackets
$[\mathbf{F}]$ denote the discontinuity of the field $\mathbf{F}$ when crossing the
surface $\Sigma$, and $\mathbf{n}$ is a unit normal to this surface usually used in
formulation of matching conditions~\cite{Stratton}.

The physical origin of the model leads to  a Debye-type cutoff $K$ on the
surface-parallel wavenumbers of waves that the fluid can support. However if one
digresses from this cutoff then the Maxwell equations (\ref{eq-7a}) and (\ref{eq-7b})
with matching conditions (\ref{eq-8}) and (\ref{eq-9}) can be regarded as a local quantum
field model that is interesting by himself. First of all the question arises here
concerning the analysis of divergences. As
known~\cite{NLS,Bordag,Vassilevich,Kirsten,Od,Ten}, to this end one has to calculate the
heat kernel coefficients for the differential operator garnering the dynamics in this
model (see Eqs.\ (\ref{eq-7a}) -- (\ref{eq-9})).

\section{Solution to the Maxwell equations for a flat plasma sheet}
\label{sec3} When constructing the normal modes in this problem it is convenient to use
the Hertz potentials~\cite{HdP1,Stratton,HdP2}. In view of  high symmetry of the flat
plasma layer the electric ($\mathbf{E}$) and magnetic ($\mathbf{B}$) fields are expressed
in terms of electric $(\bm{\Pi}')$ and magnetic $(\bm{\Pi}'')$ Hertz vectors possessing
only one nonzero component
\begin{equation}
\label{2-1} \bm{\Pi}'= \mathbf{e}_ze^{i\mathbf{ks}}\Phi(z), \quad  \bm{\Pi}''=
\mathbf{e}_ze^{i\mathbf{ks}}\Psi(z)\,{.}
\end{equation}
Here the plasma layer is taken as the $(x,y)$ coordinate plane, the axes $z$ is normal to
this plane, $\mathbf{k}$ is a two-component wave vector parallel to the plasma sheet and
$\mathbf{s}=(x,y)$; $\mathbf{e}_x,\;\mathbf{e}_y$, and $\mathbf{e}_z$ are unit base vectors
in this coordinate system. The common time-dependent factor $e^{-i\omega t}$ is dropped.

For  given Hertz vectors $\bm{\Pi}'$ and $\bm{\Pi}''$ the electric and magnetic fields are
constructed by the formulas
\begin{eqnarray}
\mathbf{E}=\bm{\nabla}\times\bm{\nabla}\times\bm{\Pi}', &\quad&
\mathbf{B}=-i\,\frac{\omega}{c}\,\bm{\nabla}\times\bm{\Pi}'\quad (\text{TM - modes},
B_z=0);\label{2-2}\\
\mathbf{E}=i\,\frac{\omega}{c}\,\bm{\nabla}\times\bm{\Pi}'',&\quad&
\mathbf{B}=\bm{\nabla}\times\bm{\nabla}\times\bm{\Pi}'' \quad (\text{TE - modes},
E_z=0)\,{.} \label{2-3}
\end{eqnarray}
Substituting here $\bm{\Pi}'$ and $\bm{\Pi}''$ by Eq.\ (\ref{2-1}) one obtains for the
TE-modes
\begin{eqnarray}
\mathbf{E}&=&(-k_y\mathbf{e}_x+k_x\mathbf{e}_y)\,\frac{\omega}{c}\,e^{i\mathbf{ks}}\Psi(z),
\label{2-4}\\
\mathbf{B}&=&i(k_x\mathbf{e}_x+k_y\mathbf{e}_y)e^{i\mathbf{ks}}\Psi'(z)+\mathbf{e}_z
k^2e^{i\mathbf{ks}} \Psi(z)\label{2-5}
\end{eqnarray}
and for the TM-modes
\begin{eqnarray}
 \mathbf{E}&=&i(k_x\mathbf{e}_x+k_y\mathbf{e}_y)e^{i\mathbf{ks}}\Phi'(z)+\mathbf{e}_z
k^2e^{i\mathbf{ks}} \Phi(z)\label{2-6} \\
\mathbf{B}&=&(k_y\mathbf{e}_x-k_x\mathbf{e}_y)\,\frac{\omega}{c}\,e^{i\mathbf{ks}}\Phi(z),
\label{2-7}
\end{eqnarray}
The fields (\ref{2-4}) -- (\ref{2-7}) will obey the Maxwell equations (\ref{eq-7a})
and(\ref{eq-7b}) outside the plasma layer $(z=0)$ when the functions
$e^{i\mathbf{ks}}\Phi(z)$ and $e^{i\mathbf{ks}}\Psi(z)$ are the eigenfunctions of the
operator $(-\Delta)$ with the eigenvalues $\omega^2/c^2$ ($\Delta$ is the
three-dimensional Laplace operator):
\begin{eqnarray}
\label{2-8} -\Delta
e^{i\mathbf{ks}}\Phi(z)&=&\frac{\omega^2}{c^2}\,e^{i\mathbf{ks}}\Phi(z)\,{,}
\\
-\Delta e^{i\mathbf{ks}}\Psi(z)&=&\frac{\omega^2}{c^2}\,e^{i\mathbf{ks}}\Psi(z)\,{.}
\label{2-9}
\end{eqnarray}
We deduce from here
\begin{eqnarray}
\label{2-10} -\Phi''(z)&=&\left (\frac{\omega^2}{c^2}\,-k^2 \right ) \Phi(z)\,{,}
\\
-\Psi''(z)&=&\left (\frac{\omega^2}{c^2}\,-k^2 \right )\Psi(z)\,{,} \label{2-11}\\
-\infty <z<\infty,&& z\neq 0, \quad 0\leq k^2<\infty\,{.} \nonumber
\end{eqnarray}

The substitution of the fields (\ref{2-4}) -- (\ref{2-7}) into the matching conditions
(\ref{eq-8}) and (\ref{eq-9}) gives the analogous conditions at the point $z=0$ for the
functions $\Phi (z)$ and $\Psi (z)$:
\begin{eqnarray}
&\left [ \Phi(0) \right]=-2q {\displaystyle \left (\frac{ c}{ \omega }\right )^2}
\Phi'(0), \quad \left [ \Phi'(0) \right]=0\,{,}&\label{2-12}
\\
&\left [ \Psi(0) \right]=0, \quad  \left [ \Psi'(0) \right]=2q  \Psi(0)\,{,}&\label{2-13}
\end{eqnarray}
where the notation is introduced
\begin{equation}
\label{2-14} \left [ F(z) \right]=F(z+0)-F(z-0)\,{.}
\end{equation}

The matching conditions (\ref{2-12}) are very specific, they involve the eigenvalues of
the whole initial spectral problem (\ref{2-8}). It implies, specifically, that  the
dynamics of TM-modes along the $z$ axes 'feels' the parallel dimensions $(x,y)$. Thus for
the TM-modes the initial spectral problem does not factorizes completely  into two
independent boundary value problems along the $z$ axis and in parallel directions.

In order to complete the formulation of the boundary spectral problems (\ref{2-10}) --
(\ref{2-13}) we have to specify the behaviour of the functions $\Phi(z)$ and $\Psi(z)$
when $|z|\to \infty$. Proceeding from the physical content of the problem in question, we
shall consider such functions  $\Phi(z)$ and $\Psi(z)$ which either oscillate
$(p^2\equiv\omega^2/c^2-k^2>0)$ or go down $(\kappa^2\equiv k^2-\omega^2/c^2>0)$ when
$|z|\to \infty$. In the first case we are dealing with the scattering states and in the
second one the solutions to the Maxwell equations describe the surface
plasmon.\footnote{In electrodynamics of continuous media the surface plasmon is a special
solution to the Maxwell equations which describes the electric and magnetic fields
exponentially decreasing in the direction normal to the interface between two material
media~\cite{Frenkel,Barton3,Ron}. The plasmon solution exists only in the TM-sector. It
is these configurations of electromagnetic field that contribute to the Casimir force
between two semi-infinite material media~\cite{Van-Kampen,Gerlach}.}

The  spectral problem for the TE-modes (\ref{2-11}) and (\ref{2-13}) with the regarding
boundary conditions is a selfadjoint spectral problem.  Substituting the TE-fields
(\ref{2-4}) and (\ref{2-5}) into the initial Maxwell equations with $\delta$-sources
(\ref{eq-2}) we arrive at the equation for $\Psi(z)$ with the $\delta$-potential
(instead of Eq.\ (\ref{2-11}))
\begin{equation}
\label{2-14a} -\Psi''(z)+2q\,\delta (z)\Psi(z) =p^2\,\Psi(z){,}\quad
\frac{\omega^2}{c^2}=k^2+p^2{.}
\end{equation}
Integration of this equation in vicinity of the point $z=0$ leads to the matching
conditions (\ref{2-13}) (see, for example, \cite{Albeverio}). The potential in Eq.\
(\ref{2-14a}) is equal to $ 2 q\,\delta(z)$. As it was mentioned above, the spectral
problem for the TM-modes (\ref{2-10}), (\ref{2-12}) possesses a peculiarity, namely, the
matching conditions (\ref{2-12}) involve the eigenvalue $\omega/c$. Obviously, the
selfadjointness condition is not satisfied here.

The scattering problem for the functions $\Phi(z)$ and $\Psi(z)$ is formulated on a line:
$-\infty <z< \infty$. As known~\cite{Barton,Newton}, the one-dimensional scattering
problem has features  in comparison with description of the scattering processes in
three-dimensional space.

The scattering states for the TE-modes are described by the functions
\begin{eqnarray}
\Psi(z)&=&C_1e^{ipz}+C_2e^{-ipz},\quad z<0\,{,} \label{2-15}
\\
\Psi(z)&=&C_3e^{ipz}, \quad z>0,\quad p>0, \quad p^2=\frac{\omega^2}{c^2}-k^2\,{.}
\label{2-16}
\end{eqnarray}
The initial wave is coming from the left. Matching conditions (\ref{2-13})  give the
following relation between the constants $C_i\quad (i=1,2,3)$:
\begin{eqnarray}
\label{2-18} C_1+C_2&=&C_3\,{,}
\\
ip\,(C_3-C_1+C_2)&=&2q\,C_3\,{.} \label{2-19}
\end{eqnarray}
From here we obtain for the reflection and transmission coefficients
\begin{equation}
\label{2-20}
{\cal{R}}^{\text{TE}}\equiv\frac{C_2}{C_1}=\frac{-iq}{p+iq}=i\sin\eta\,e^{i\eta},\quad
{\cal{T}}^{\text{TE}}=\frac{p}{p+iq}=\cos\eta\,e^{i\eta}\,{,}
\end{equation}
where $\eta(p)$ is the phase shift for the scattering of the
TE-modes. When initial waves is coming from the right we obtain
the same reflection and transition coefficients. There is no other
conditions on $p$ besides the positivity $p>0$ (continuous
spectrum). The scattering matrix, $S(p)$, is determined by the
phase shift ~$\eta(p)$
\begin{equation}
\label{2-20a} \tan\eta (p) =-\frac{p}{q},\quad p>0
\end{equation}
in a standard way
\begin{equation}
\label{2-20b} S(p)=e^{2i\,\eta(p)}{.}
\end{equation}

Between TE-modes there is no solutions which go down when $|z|\to \infty$. Indeed, such
solutions, in accord with Eq.\ (\ref{2-11}) would have the form
\begin{eqnarray}
\Psi(z)&=&C_1 e^{\kappa z}, \quad z <0, \nonumber
\\
\Psi(z)&=&C_2 e^{-\kappa z},\quad z>0,\quad \kappa
=+\sqrt{k^2-\frac{\omega^2}{c^2}}>0\,{.}\label{2-21}
\end{eqnarray}
Substitution of Eq.\  (\ref{2-21}) into the  matching conditions (\ref{2-13}) gives
\begin{equation}
\label{2-22} C_1=C_2 \quad \text{and} \quad \kappa =-q<0\,{.}
\end{equation}
Thus, there is no solutions of the form (\ref{2-21}) with positive~$\kappa$. This is in
accord with the known fact~\cite{Albeverio} that the potential defined by the Dirac
$\delta$-function $\lambda\,\delta (z)$ leads to bound state only for $\lambda <0$. In
our case $\lambda =2q>0$ (see Eq.\ (\ref{2-14a})).

 Finally, the
TE-modes have the spectrum
\begin{equation}
\label{2-23} \frac{\omega^2(\mathbf{k},p)}{c^2}=k^2+p^2,\quad \mathbf{k}\in
\mathbb{R}^2,\quad 0\leq p<\infty\,{.}
\end{equation}
Here the contribution $k^2$ is given by the free waves propagating in directions parallel
to the plasma layer, and the contribution $p^2$ corresponds to the one-dimensional
scattering in normal direction with the phase shift~(\ref{2-20a}).

Proceeding in the same way one obtains  for the scattering of the TM-modes:
\begin{equation}
\label{2-24} {\cal R}^{TM}=\frac{ip\,q}{p^2+k^2+ip\,q}=-i\sin \mu\,e^{i\mu}, \quad {\cal
T}^{\text{TM}}=\frac{p^2+k^2}{p^2+k^2+ip\,q}=\cos\mu\,e^{i\mu}\,{,}
\end{equation}
where the phase shift $\mu(p,k)$ is defined by
\begin{equation}
\label{2-25} \tan \mu(p,k)=-\frac{p\,q}{p^2+k^2}, \quad p\geq 0\,{.}
\end{equation}
The scattering matrix is defined here by the formula analogous to Eq.\ (\ref{2-20b}):
$S(p,k)=e^{2i\mu(p,k)}$.

It turns out that between TM-modes there are solutions which not only oscillate when
$|z|\to \infty$, but also go down in this limit. In this case the function $\Phi (z)$ is
defined by the equation
\begin{equation}
\label{2-26} \Phi''(z) -\left ( k^2-\frac{\omega^2}{c^2} \right )\Phi(z)=0\,{,}
\end{equation}
where
\begin{equation}
\label{2-27} k^2-\frac{\omega^2}{c^2}\equiv\kappa^2>0\,{.}
\end{equation}
The solution we are interested in should have the form
\begin{eqnarray}
\Phi(z)&=&C_1e^{\kappa z}, \quad z<0,\nonumber
\\
\Phi(z)&=&C_2e^{-\kappa z}, \quad z>0, \label{2-28}
\end{eqnarray}
where $\kappa=+\sqrt{k^2-\omega^2/c^2}>0$. The matching conditions (\ref{2-12}) give rise
to the equation  for~$\kappa$:
\begin{equation}
\label{2-29} \kappa^2+\kappa \, q-k^2=0\,{,}
\end{equation}
the positive root of which is
\begin{equation}
\label{2-30} \kappa=\sqrt{\frac{q^2}{4}+k^2}-\frac{q}{2}\,{.}
\end{equation}
For the respective frequency squared we derive by making use of Eq.\ (\ref{2-27})
\begin{equation}
\label{2-31} \frac{\omega^2_{\text{sp}}}{c^2}=\frac{q}{2} \left( \sqrt{q^2+4k^2}-q \right
) \geq 0{.}
\end{equation}
Thus, the frequency of the surface plasmon is real, and the
solution obtained oscillates in time instead of damping.

Finally, the spectrum of the TM-modes in the problem under study have two branches: i)
the photon branch and ii) the surface plasmon branch. The photon spectrum of TM-modes is
defined by
\begin{equation}
\label{2-31a} \frac{\omega_{\text{ph}}^2(\mathbf{k},p)}{c^2}=\mathbf{k}^2+p^2,\quad
\mathbf{k}\in\mathbb{R}^2, \quad 0\leq p<\infty\,{.}
\end{equation}
Here the contribution $\mathbf{k}^2$, as in Eq.\ (\ref{2-23}) is given by the free waves
propagating in directions parallel to the plasma layer and $p^2$ is the contribution of
the one-dimensional scattering with the phase shift~(\ref{2-25}) in normal, to the plasma
sheet, direction. The surface plasmon branch of the spectrum is described by
\begin{equation}
\label{2-32} \frac{\omega_{\text{sp}}^2(\mathbf{k})}{c^2}=\frac{q}{2}\left (
\sqrt{q^2+4k^2}-q \right), \quad \mathbf{k}\in\mathbb{R}^2\,{.}
\end{equation}

 In the spectral problem (\ref{2-10}), (\ref{2-12}) with negative  $q$ the parameter $\kappa$
defined by Eq.\ (\ref{2-30}) is positive, however the respective frequency  squared  is
negative
\begin{equation}
\label{2-33} \frac{\omega^2_{q<0}}{c^2}=-\frac{1}{2}\left ( q^2+|q|\sqrt{q^2+4k^2} \right
){.}
\end{equation}
Hence, in this case we have a resonance state instead of the bound state (surface plasmon).

\section{Spectral functions in a flat plasma sheet model}
\label{sec4} In the previous section  we have determined the spectrum in the model under
consideration (see Eqs.\ (\ref{2-23}), (\ref{2-31a}), and (\ref{2-32})) therefore we can
proceed to construction of the spectral functions in this model, namely, the spectral
zeta function
\begin{equation}
\label{4-1} \zeta(s)= \text{Tr}\; L^{-s}=\sum_n\lambda _n^{-s}
\end{equation}
and (integrated) heat kernel
\begin{equation}
\label{4-2} K(t)=\text{Tr}\; (e^{-t\,L})=\sum_n e^{-\lambda _n t}\,{,}
\end{equation}
where $L$ is the differential operator in Eqs.\ (\ref{2-10}) and (\ref{2-11}) with
matching conditions at $z=0$ (\ref{2-12}), (\ref{2-13}) and conditions specified above
for $|z|\to \infty$; $\lambda_n$ are the eigenvalues given by Eqs.\ (\ref{2-23}),
(\ref{2-31a}), and (\ref{2-32}). In Eq.\ (\ref{4-2}) $t$ is an auxiliary variable, $0\leq
t < \infty $. In our case $L=-\Delta$ and $t$ has the dimension [length]$^2$.

 As a preliminary, it is worth
stipulating the procedure of spectral summation  entered Eqs. (\ref{4-1}) and
(\ref{4-2}). Obviously the contribution to the spectral functions generated by free waves
is given by a simple integration over the respective wave vector $\mathbf{k}$:
\begin{equation}
\label{4-3} \int \frac{d^2\mathbf{k}}{(2\pi)^2}\cdots = \int_0^\infty \frac{k\,dk}{2\pi}
\cdots\;{.}
\end{equation}

The contribution of the scattering states we define by making use of the phase shift
method, namely, the integration over $dp$ will be carried out with the spectral density
$\rho(p)$ (the function of spectral shift):
\begin{equation}
\label{4-4} \rho(p) =\frac{1}{2\pi i}\,\frac{d}{dp}\,\ln {S(p)} =
\frac{1}{\pi}\,\frac{d}{dp}\, \delta (p)\,{,}
\end{equation}
where $\delta(p)$ is the phase shits for TE and TM modes given by Eqs.\ (\ref{2-20a}) and
(\ref{2-25}), respectively.

\subsection{TE modes}Taking into account all this we get for the spectral zeta function in the
TE-sector of the model
\begin{equation}
\label{4-5} \zeta^{\text{TE}}(s)=\int \frac{d^2\mathbf{k}}{(2\pi)^2}\int_0^\infty dp\,
(k^2+p^2)^{-s}\frac{1}{\pi}\,\frac{d}{dp}\,\eta(p)=\frac{q}{2\pi^2}\int^\infty
_0\frac{dp}{p^2+q^2}\int_0^\infty\frac{k\,dk}{(k^2+p^2)^s}\,{.}
\end{equation}
The left-hand side  of Eq.\ (\ref{4-5}) is defined in the semi-plane $\text{Re}\;s>1$ and
this representation can be analytically extended all over the complex plane $s$ save for
separate points. In order to do this it is sufficient to express the integrals in Eq.\
(\ref{4-5}) in terms of the gamma-function
\begin{equation}
\zeta^{\text{TE}}(s)=\frac{q^{2-2s}}{8\pi^2}\,\frac{\Gamma(3/2-s)
\,\Gamma(s-1/2)\,\Gamma(s-1)}{\Gamma(s)}=\frac{q^{2-2s}}{8\pi(1-s)\cos(\pi s)}\,{.}
\end{equation}

The integrated heat kernel for TE-modes is given by
\begin{equation}
\label{4-6} K^{\text{TE}}(t)=K_0^{(d=2)}(t)\cdot K^{(d=1)}(t)\,{,}
\end{equation}
where $K_0^{(d=2)}(t)$ is the free two-dimensional heat kernel responsible for free waves
propagating in directions parallel to the plasma layer
\begin{equation}
\label{4-7} K_0^{(d=2)}(t)=\int_0^\infty \frac{k\,dk}{2\pi}\,e^{-k^2t}=\frac{1}{4\pi t
}\,{,}
\end{equation}
and
\begin{eqnarray}
\label{4-8} K^{(d=1)}(t)&=&\frac{1}{\pi} \int_0^\infty dp\,e^{-p^2t}\frac{d}{dp}
\arctan\left (-\frac{q}{p} \right)=\frac{q}{\pi}\int_0^\infty
\frac{e^{-p^2t}}{p^2+q^2}\,dp \nonumber \\
&=&  \frac{1}{2}\,e^{tq^2}\,\left [1- \text{erf}\left (q\sqrt{t}\right )\right ]
=\frac{1}{2}\,e^{tq^2}\,\text{erfc}\left (q\sqrt{t}\right ){,}
\end{eqnarray}
where $\text{erf}\left (q\sqrt{t}\right )$ is the probability integral~\cite{GR,AS}.

In the plasma sheet model under consideration the parameter $q$ is strictly positive.
However  the spectral problem (\ref{2-14a}) can be considered  for the negative $q$ also
when it possesses a bound state~\cite{Albeverio}. The contribution of this state should be
taken into account when calculating the respective heat kernel $K_{q<0}(t)$. But this heat
kernel can be obtained by analytical continuation of Eq.\ (\ref{4-8})
\begin{equation}
 K_{q<0}(t)=\frac{1}{2}\,e^{tq^2}\left [ 1-\text{erf}\left (-|q|\sqrt{t}\right )\right]=
\frac{1}{2}\,e^{tq^2}\left [ 1+\text{erf}\left (|q|\sqrt{t}\right ) \right]{.} \label{4-8a}
\end{equation}
Comparison  Eqs.\ (\ref{4-8})  and (\ref{4-8a}) gives the following simple relation between
the integrated heat kernel for $\delta$ potentials with positive and negative coupling
constants
\begin{equation}
\label{4-8b} K_{q>0}(t)+ K_{q<0}(t)=e^{tq^2}\,{,}
\end{equation}
where  $K_{q>0}(t)$ is given by Eq.\ (\ref{4-8}).

The structure of divergences in quantum field theory is determined by the coefficients of
the asymptotic expansion of the heat kernel when $t\to +0$
\begin{equation}
\label{4-9} K(t)= (4\pi t)^{-d/2}\sum _{n=0,1,2,\ldots} t^{n/2}B_{n/2}+\text{ES}\,{,}
\end{equation}
where $d$ is the dimension of the configuration space in the problem at hand, ES stands
for the exponentially small corrections. The coefficients $B_{n/2}$ for TE-modes can be
derived in two different ways: i) by making use of Eqs.\ (\ref{4-6}) -- (\ref{4-8}) which
determine the heat kernel explicitly and ii) by applying the known relation between the
residua of the product $\zeta (s)\, \Gamma (s)$ and $B_{n/2}$~:
\begin{equation}
\label{4-10} \frac{B_{n/2}}{(4\pi)^{d/2}}=\lim _{s\to\frac{d-n}{2}}\left (
s+\frac{n-d}{2} \right ) \Gamma(s)\, \zeta (s), \quad n=0,1,2,\ldots \,{.}
\end{equation}

 Both calculation schemes give the same values for the coefficients $B_{n/2}$ with $n\neq 0$~:
\begin{eqnarray}
&B_{1/2}=\sqrt{\pi}, \quad B_1=-2q, \quad B_{3/2}=\sqrt{\pi} q^2,\quad
B_2=-\frac{4}{3}\,q^3,&\nonumber \\
&B_{5/2}=\frac{\sqrt{\pi}}{2}\,q^4,\quad B_3=-\frac{8}{15}\,q^5, \quad
B_{7/2}=\frac{\sqrt{\pi}}{6}\,q^6, \quad \ldots \,{.}&
\end{eqnarray}
Equations (\ref{4-6}) -- (\ref{4-8}) give for the coefficient $B_0$
\begin{equation}
\label{4-12} B_0=0\,{,}
\end{equation}
while employment of the zeta  function (\ref{4-5}) and relation (\ref{4-10}) leads to the
nonzero value
\begin{equation}
\label{4-13} B_0=-\frac{1}{q}\,{.}
\end{equation}

The different values of $B_0$ in Eqs.\ (\ref{4-12}) and (\ref{4-13}) are the consequence
of the noncompact configuration space, ${\mathbb R}^3$, in the problem at hand. In the
case of compact manifolds the coefficient $B_0$ is equal to the volume $V$ of the
manifold: $B_0=V$. Definition of $B_0$ for a noncompact configuration space requires a
special consideration~\cite{NPD,SW}. The correct value of $B_0$ in our case is formula
(\ref{4-12}) derived from the explicit heat kernel (\ref{4-6}) -- (\ref{4-8}). The point
is,  the contribution  of unbounded space $\mathbb{R}^3$ is subtracted from the spectral
density (\ref{4-4}). Hence this contribution is subtracted also from the multiplier
$K^{(d-1)}(t)$ in Eq.\ (\ref{4-6}) and, due to the factorization  this formula, the
contribution of $\mathbb{R}^3$ is automatically subtracted  from the total kernel $K(t)$.

When calculating the zeta function $\zeta^{\text{TE}}(s)$ the spectral density
(\ref{4-4}) with subtracted $\mathbb{R}^3$-contribution has been used also. However in
Eq.\ (\ref{4-5}), unlike Eq.\ (\ref{4-6}), there is no factorization of the contributions
of waves propagating in directions parallel to the plasma plane and in normal to this
plane direction. It is this point that lead to nonzero value of $B_0$ in Eq.\
(\ref{4-13}).

\subsection{TM-modes}
In the case of TM-modes there are two branches of spectrum: i) the photon branch
(\ref{2-31a}) and ii) the surface plasmon branch (\ref{2-32}).

By making use of Eqs.\ (\ref{2-25}) and (\ref{4-4}) we obtain for the photon zeta
function in the TM-sector of the model
\begin{eqnarray}
\zeta^{\text{TM}}_{\text{ph}}(s)&=&\int\frac{d^2\mathbf{k}}{(2\pi)^2}\int_0^\infty dp\,
(k^2+p^2)^{-s}\frac{1}{\pi}\,\frac{d}{dp}\,\mu(p,k)\nonumber
\\&=&\frac{q}{2\pi^2}\int_0^\infty
k\,dk\int_0^\infty\frac{dp}{(k^2+p^2)^s}\,\frac{p^2-k^2}{(k^2+p^2)^2+q^2p^2}\,{.}\label{4-14}
\end{eqnarray}
The integration in (\ref{4-14}) can be easily done by introducing first the polar
coordinates
\begin{eqnarray}
&\label{4-15} k=r\sin \varphi,\quad p=r\cos\varphi,\quad dk\,dp=r\,dr\,d\varphi,\quad
&\nonumber \\&0\leq \varphi\leq \pi/2,\quad 0\leq r\leq \infty\,{.}&
\end{eqnarray}
In terms of new variables formula (\ref{4-14}) acquires the form
\begin{equation}
\label{4-16}
\zeta^{\text{TM}}_{\text{ph}}(s)=\frac{q}{2\pi^2}\int_0^{\pi/2}d\varphi\int_0^\infty dr
\frac{r^{2-2s}\sin \varphi\, (2\cos^2\varphi-1)}{r^2+q^2\cos^2\varphi}\,{.}
\end{equation}
Now it is natural to make the following substitution of the variables \begin{equation}
\label{4-16a} \cos \varphi =x,\quad -\sin \varphi\,d\varphi =dx\,{.} \end{equation}
 As a
result one gets
\begin{equation}
\label{4-17} \zeta^{\text{TM}}_{\text{ph}}(s)=\frac{q}{2\pi^2}\int_0^1dx\,(2x^2-1)
\int_0^\infty\frac{r^{2-2s}dr}{r^2+q^2x^2}\,{.}
\end{equation}
The integral over $r$ in (\ref{4-17}) exists in the region
\begin{equation}
\label{4-18} 1/2<\text{ Re }s<3/2\,{.}
\end{equation}
and the subsequent integration over $x$ can be done for
\begin{equation}
\label{4-19} 1/2<\text{ Re }s<1\,{.}
\end{equation}
Thus the integral representation (\ref{4-17}) can be analytically extended all over the
complex $s$ plane. This yields
\begin{equation}
\label{4-20}
\zeta^{\text{TM}}_{\text{ph}}(s)=\frac{q^{2-2s}}{8\pi^2}\frac{s}{(s-1)(2-s)}\,\Gamma(3/2-s)
\,\Gamma(s-1/2) =\frac{q^{2-2s}}{8\pi}\frac{s}{(1-s)(2-s)\cos{\pi s}}\,{.}
\end{equation}

 Let us proceed to the construction of the heat kernel for the photon branch of the
spectrum in the TM-sector of the model at hand. By making use of Eqs.\ (\ref{2-31a}),
(\ref{4-2}),(\ref{4-3}), and (\ref{4-4}), we obtain
\begin{equation}
\label{4-21} K^{\text{TM}}_{\text{ph}}(t) =\frac{q}{2\pi^2}\int_0^\infty k\,dk
\,e^{-k^2t}\int_0^\infty dp\, e^{-p^2t}\frac{p^2-k^2}{(k^2+p^2)^2+q^2p^2}\,{.}
\end{equation}
Further we again  introduce  the polar coordinates (\ref{4-15}) and then make the change
of variables (\ref{4-16a}). Finally calculations are reduced to the substitution of the
term $r^{-2s}$ in Eq.\ (\ref{4-17}) by $e^{-r^2t}$:
\begin{equation}
\label{4-22} K^{\text{TM}}_{\text{ph}}(t) =\frac{q}{2\pi^2}\int_0^1
dx\,(2x^2-1)\,I(x)\,{,}
\end{equation}
where
\begin{eqnarray}
I(x)&=& \int_0^\infty\frac{r^2e^{-r^2t}}{r^2+q^2x^2}\,dr=\frac{1}{2}\sqrt{\frac{\pi}{t}}
-\frac{\pi q x}{2}e^{q^2x^2t}\left [1-\text{erf} \left (qx\sqrt t\right )\right ] \nonumber \\
&=&\frac{1}{2}\sqrt{\frac{\pi }{t}}-\frac{\pi q x}{2}\left [
e^{q^2x^2t}-\frac{2}{\sqrt{\pi}}\sum_{k=0}^\infty \frac{2^k(q x
\sqrt{t})^{2k+1}}{(2k+1)!!}\right]{.} \label{4-23}
\end{eqnarray}
We have used here the series representation for the function $\text{erf }(qx\sqrt
t)$~\cite{GR,AS}.

Equations (\ref{4-10}), (\ref{4-20}), on the one hand, and  the explicit form of the heat
kernel, Eqs.\  (\ref{4-22}) and (\ref{4-23}), on the other hand give the same value for
the coefficients $B_{n/2}$ in expansion (\ref{4-9}) for all $n$ save for $n=0,1$:
\begin{eqnarray} &B_1=-\frac{2}{3}\,q,\quad B_{3/2}=0,\quad B_2=\frac{4}{15}\,q^3,& \nonumber
\\
& B_{5/2}=-\frac{\sqrt{\pi}}{6}\,q^4, \quad B_3=\frac{8}{35}\,q^5,\quad
B_{7/2}=-\frac{\sqrt{\pi}}{12}\,q^6\,{.}\quad&\label{4-24}
\end{eqnarray}
The explicit form of the integrated heat kernel in question (\ref{4-22}) and (\ref{4-23})
shows that two first coefficients in expansion (\ref{4-9}) vanish
while formulas (\ref{4-22}) and  (\ref{4-23}) give
\begin{equation}
\label{4-26} B_0=-\,\frac{3}{q},\quad B_{1/2}=\sqrt{\pi}\,{,}
\end{equation}
This discrepancy implies that calculation of the first heat kernel coefficients through
the respective spectral zeta function requires a special consideration in the  case of
unbounded configuration space.

Let us address now the spectral functions for the spectrum branch (\ref{2-32}) generated
by the surface plasmon. Equations (\ref{2-32}), (\ref{4-1}), and (\ref{4-3}) give
\begin{equation}
\label{4-27} \zeta^{\text{TM}}_{\text{sp}}(s)=\int\frac{d^2\mathbf{k}}{(2\pi)^2}\,
\frac{\omega^{-2s}_{\text{sp}}(\mathbf{k})}{c^{-2s}} =\left(\frac{2}{q} \right)^s \int
_0^\infty \frac{k\,dk}{2\pi}\left( \sqrt{q^2+4k^2}-q\right )^{-s}{.}
\end{equation}
The convergence of this integral in the region $k\to 0$ requires $\text{Re }s<1$, but
when $k\to \infty$ it exists only if $\text{Re }s>2$. Thus for the spectrum (\ref{2-32})
it is impossible to construct the zeta function by making use of the analytical
continuation method, since one cannot define this function in any finite domain of the
complex plane~$s$.

  However it turns out that for this branch of the spectrum the heat kernel can be
constructed explicitly. Indeed, by making use of Eqs.\ (\ref{4-2}), (\ref{4-3}), and
(\ref{2-32}) we obtain
\begin{equation}
\label{4-28} K^{\text{TM}}_{\text{sp}}(t)=\int\frac{d^2\mathbf{k}}{(2\pi)^2}\exp\left[
-\frac{\omega^2_{\text{sp}}(\mathbf k)}{c^2}\,t \right]=\int_0^\infty
\frac{k\,dk}{2\pi}\exp\left[ -\frac{q}{2}\left ( \sqrt{q^2+4k^2}-q \right)t \right]{.}
\end{equation}
The change of variables
\[
\frac{q}{2}\left ( \sqrt{q^2+4k^2}-q \right)=x, \quad k\,dk =\left (
\frac{x}{q^2}+\frac{1}{2} \right )dx
\]
reduces the integral (\ref{4-28}) to the form
\begin{equation}
\label{4-29} K^{\text{TM}}_{\text{sp}}(t)=\int_0^\infty\frac{dx}{2\pi}\left(
\frac{x}{q^2}+\frac{1}{2} \right )e^{-tx}=\frac{1}{2\pi q^2t^2}+\frac{1}{4\pi t} \,{.}
\end{equation}
The first term in Eq.\ (\ref{4-29}) is absent in the standard expansion (\ref{4-9}), and
the second term in (\ref{4-29}) yields $B_{1/2}=2\sqrt \pi$. The rest of coefficients
$B_{n/2}$ with $n\neq 1$ equal zero. Thus the  surface plasmon with the spectrum
(\ref{2-32}) is a simple, and at the same time of a physical meaning, model that has no
spectral zeta function, but the the respective (integrated) heat kernel exists  though with
a nonstandard asymptotic expansion.  It is not surprising because we are dealing here with
a singular point interaction described by the matching conditions (\ref{2-12}).

\section{Local heat kernel}
\label{sec5} In the TE-sector of the model under consideration one can construct, in an
explicit form, the local heat kernel or Green function
\begin{equation}
\label{5-1} K(\mathbf{r},\mathbf{r'};t)=<\mathbf{r}|e^{-t\,L}|\mathbf{r'}> =\sum_n
\varphi^*_n(\mathbf{r})\varphi_n(\mathbf{r'})e^{-\lambda_n t}\,{.}
\end{equation}
We are using here the same notations as in Eq.\ (\ref{4-2}) and $\varphi_n (\mathbf{r})$
are normalized eigenfunctions in the spectral problems at hand
\begin{equation}
\label{5-1a} L\varphi_n(\mathbf{r'})=\lambda_n \varphi_n(\mathbf{r'}) \quad
L=-\Delta\,{.}
\end{equation}
 From the definition
(\ref{5-1}) it follows, in particular, that the local heat kernel obeys the heat
conduction equations with respect the variables ($\mathbf{r},t $) and ($\mathbf{r}',t $)
\begin{eqnarray}
 \left ( \Delta_{\mathbf{r}}-\frac{\partial}{\partial t}  \right
)K(\mathbf{r},\mathbf{r}';t)&=&0\,{,}\label{5-1b} \\ \left (
\Delta_{\mathbf{r}'}-\frac{\partial}{\partial t}  \right )K(\mathbf{r},\mathbf{r}';t)&=&0
\label{5-1c}\end{eqnarray} and initial condition
\begin{equation}
\label{5-1d} K(\mathbf{r},\mathbf{r}';t)\to \delta(\mathbf{r}-\mathbf{r}'), \text{ when }
t\to 0^{+}.
\end{equation}

Obviously in the case of TE-modes $K(\mathbf{r},\mathbf{r'};t)$ can be represented in a
factorized form
\begin{equation}
\label{5-2} K(\mathbf{r},\mathbf{r'};t)=K_0^{(d=2)}(\mathbf{s},\mathbf{s'};t)\cdot
K(z,z';t),\quad \mathbf{r}=(\mathbf{s}, z),\quad \mathbf{s}=(x,y)\,{,}
\end{equation}
where $K_0^{(d=2)}(\mathbf{s},\mathbf{s'};t)$ is the free heat kernel in directions
parallel to the plane $z=0$:
\begin{equation}
\label{5-3} K_0^{(d=2)}(\mathbf{s},\mathbf{s'};t)=\frac{1}{4\pi t}\,\exp\left
[-\frac{(\mathbf{s}-\mathbf{s'})^2}{4t}\right ]\,{.}
\end{equation}

For constructing the heat kernel $K(z,z';t)$ along the infinite $z$-axes we shall use the
technique of  integral equations developed in our previous paper~\cite{PNB}. These
equations naturally arise when representing the Green function to be found in terms of
heat potentials of simple and double layers. With respect of its first argument $z$ the
heat kernel $K(z,z';t)$ obeys the matching conditions (\ref{2-13}) and when $t\to 0^+$ it
satisfies the initial condition (\ref{5-1d}).

 In what follows it is convenient to represent the heat kernel we are looking
for in terms of four components depending on the range of its arguments
\begin{equation} \label{5-4} K(z,z';t)=
\begin{cases}
K_{-+}(z,z';t), &z<0,\;z'>0\,{,}\\
K_{++}(z,z';t),  & z,\,z'>0\,{,} \\
K_{+-}(z,z';t), &z>0,\;z'<0\,{,}\\
K_{--}(z,z';t),  &z,\,z'<0\,{.}
\end{cases}
\end{equation}

We represent the heat kernel $K(z,z';t)$ in terms of heat
potentials of a simple layers~\cite{PNB}
\begin{eqnarray}
K_{-+}(z,z';t)&=&\int_0^t d\tau\, K_0(z,0;t-\tau)\,\alpha_1
(\tau;z')\,{,} \quad z<0,\;z'>0\,{,} \label{5-5}
\\
K_{++}(z,z';t)&=&K_0(z,z';t)+\int_0^t
d\tau\,K_0(z,0;t-\tau)\,\alpha_2 (\tau;z')\,{,} \quad
z,\;z'>0\,{,} \label{5-6}
\end{eqnarray}
where $\alpha_1(\tau;z')$ and $\alpha_2(\tau;z')$ are the densities of the heat
potentials to be found and $K_0(z,z';t)$ is the free heat potential (Green's function) on
an infinite line
\begin{equation}
\label{5-7} K_0(z,z';t)=\frac{1}{\sqrt{4\pi t}}\exp\left [-\frac{(z-z')^2}{4t}\right ]{.}
\end{equation}
Substituting Eqs.\ (\ref{5-5}) and (\ref{5-6}) in matching conditions (\ref{2-13}) we
obtain
\begin{eqnarray} \int_0^t d\tau\,
K_0(0,0;t-\tau)[\alpha_1(\tau;z')-\alpha_2(\tau;z')]&=&K_0(0,z';t),\quad z'>0\,{,}
\label{5-8}
\\
2\frac{\partial K_0}{\partial z}(z=0,z';t)-4q \!\!\int_0^t \!\!\! d\tau
K_0(0,0;t-\tau)\alpha_1(\tau;z')&=&\alpha_1(\tau;z')+\alpha_2(\tau;z'){,}\;z'>0{.}
\label{5-9}
\end{eqnarray}
When deriving Eq.\ (\ref{5-9}) we have taken into account that the derivative of the
single layer potential has a jump at the interface, namely, let $V(z;t)$ is the heat
potential of a single layer with the density $\nu(\tau)$:
\begin{equation}
\label{5-10} V(z;t)=\int_0^t d\tau\, K_0(z,0;t-\tau)\,\nu(\tau)\,{,}
\end{equation}
then one can easily show that~\cite{Smirnov,PNB}
\begin{equation}
\label{5-11}  \frac{\partial V}{\partial z}\,(z=0^+;t) =-\frac{1}{2}\,\nu(t)\,{,}\quad
\frac{\partial V}{\partial z}\,(z=0^-;t) =\frac{1}{2}\,\nu(t)\,{.}
\end{equation}

On substituting in Eq.\  (\ref{5-8}) the explicit form of the free heat kernel
(\ref{5-7}) we arrive at the Abel integral equation the exact solution of which is
known~\cite{SmirnovII,WW} or it  can be derived by the Laplace transform
\begin{equation}
\label{5-12} \alpha_1(t;z)- \alpha_2(t;z)=\frac{z}{t\sqrt{4 \pi t}}\exp\left(
-\frac{z^2}{4t} \right){,}\quad z>0\,{.}
\end{equation}
In what follows it is convenient to apply to Eqs.\ (\ref{5-9}) and (\ref{5-12}) the
Laplace transform
\begin{equation}
\label{5-13} \bar f(p) =\int_0^\infty  e^{-p\,t}f(t)\,dt\,{.}
\end{equation}
As a result we obtain the set of two linear  algebraic equations for the functions $\bar
\alpha_1(p;z')$ and $\bar \alpha_2(p;z')$:
\begin{eqnarray}
\bar \alpha_1(p;z')-\bar \alpha_2(p;z') &=& e^{-z'\sqrt p}{,} \quad z'>0\,{,}
\label{5-14} \\
\bar \alpha_1(p;z')+\bar \alpha_2(p;z') &=& e^{-z'\sqrt p}-\frac{2q}{\sqrt p}\,\bar
\alpha_1(p;z'){,} \quad z'>0\,{.} \label{5-15}
\end{eqnarray}
From here we deduce
\begin{equation}
\label{5-16} \bar \alpha_1(p;z')=\frac{\sqrt p}{q+\sqrt p}\,e^{-z'\sqrt p}{,}\quad
 \bar \alpha_2(p;z')=-\frac{q}{q+\sqrt p}\,e^{-z'\sqrt p},\quad z'>0\,{.}
\end{equation}
By making use of the Laplace transform (\ref{5-13}) the initial formulas (\ref{5-5}) and
(\ref{5-6}) can be rewritten in the form
\begin{eqnarray}
\bar K_{-+}(z,z';p) &=&\frac{1}{2\sqrt p}\,e^{-|z|\sqrt p}\,\bar \alpha_1(p;z'){,}\quad
z<0,\; z'>0\,{,}
\label{5-17} \\
\bar K_{++}(z,z';p)&=&\frac{1}{2\sqrt p}\,e^{-|z-z'|\sqrt p}+\frac{1}{2\sqrt
p}\,e^{-z\sqrt p}\,\bar \alpha_2(p;z'){,}\quad z,\;z'>0\,{.} \label{5-18}
\end{eqnarray}
Substitution of Eq.\ (\ref{5-16}) into Eqs.\ (\ref{5-17}) and (\ref{5-18}) yields
\begin{eqnarray}
\bar K_{-+}(z,z';p) &=&\frac{1}{2(q+\sqrt p)}\,e^{-(|z|+z')\sqrt p},\quad z<0,\;z'>0\,{,}
\label{5-19} \\
\bar K_{++}(z,z';p) &=&\frac{1}{2\sqrt p}\,e^{-|z-z'|\sqrt p}- \frac{q}{2\sqrt
p\,(q+\sqrt p)}\,e^{-(z+z')\sqrt p},\quad z,\;z'>0\,{.} \label{5-20}
\end{eqnarray}
The inverse Laplace transform~\cite{AS,Erdelyi} applied to Eqs.\ (\ref{5-19}) and
(\ref{5-20}) gives
\begin{eqnarray}
K_{-+}(z,z';t) &=&K_0(|z|+z',0;t)-\frac{q}{2}\,e^{q(|z|+z')+q^2t}\,\text{erfc}\left (
q\sqrt t+\frac{|z|+z'}{2\sqrt t} \right ) {,}\nonumber \\
&&z<0,\,z'>0\,{,}
 \label{5-21}\\
K_{++}(z,z';t)&=&K_0(z,z';t)-\frac{q}{2}\,e^{q(z+z')+q^2t}\,\text{erfc}\left ( q\sqrt
t+\frac{z+z'}{2\sqrt t} \right ){,}\quad z,\;z'>0\,{.}
 \label{5-22}
 \end{eqnarray}

 By making use of the same technique one can derive for the components $K_{+-}(z,z';t)$
and $K_{--}(z,z';t)$ the formulas which are analogous to Eqs.\ (\ref{5-21}) and
(\ref{5-22}). It turns out that all four components of the heat kernel $K(z,z';t)$ (see
Eq.\ (\ref{5-4})) can be represented in a unique way, namely:
\begin{eqnarray}
\label{5-23} K(z,z';t)&=&K_0(z,z';t)-\frac{q}{2}\,e^{q(|z|+|z'|)+q^2t}\,\text{erfc}\left
( q\sqrt t+\frac{|z|+|z'|}{2\sqrt t} \right ){,}\nonumber \\ && -\infty <z,\;z'<
\infty,\quad z\neq 0,\;z'\neq 0{.}
\end{eqnarray}
In literature~\cite{FA,Schroedinger} the fundamental solutions to the heat conduction
equation or to the Schr\"odinger equation with the Dirac $\delta $ potential are
represented, as a rule, in an integral form
\begin{equation}
\label{5-24} K(z,z';t)=K_0(z,z';t)-q\int_0^\infty e^{-qu}\,K_0(|z|+|z'|+u,0;t)\,du\,{.}
\end{equation}
It is worth noting that the formulas (\ref{5-23}) and (\ref{5-24}) have the sense also at
the points $z=0$ and $z'=0$.

Thus the technique of integral equations developed on the basis of heat potentials in
Ref.~\cite{PNB} affords a regular and simple method for deriving, in an explicit form, the
heat kernel in the problem under consideration. In Ref.\ \cite{Solod} the representation
(\ref{5-23}) was obtained in  the framework of another approach to this problem.

 Proceeding from Eq.\ (\ref{5-24}) one can easily obtain the traced heat kernel (\ref{4-8})
by evaluating the integral
\begin{eqnarray}
K(t) &=&\int_{-\infty}^\infty dz\left [ K(z,z;t)-K_0(z,z;t) \right ]
\nonumber \\
&=&-q\int_{-\infty}^\infty dz\int_0^\infty du\,e^{-qu}\,K_0(2|z|+u,0;t)\,{.}
\end{eqnarray}
Substitution to this equation of the exact form of the free heat kernel (\ref{5-7}) gives
\begin{eqnarray}
K(t)&=& -\frac{q}{\sqrt{\pi t}}\int_0^\infty dz\int_0^\infty du\,e^{-qu}\,\exp\left[
-\frac{(2z+u)^2}{4t} \right]\nonumber \\
&=&-\frac{q}{2}\int_0^\infty du\,e^{-qu}\,\text{erfc}\left (\frac{u}{2\sqrt t}\right )
{.}
\end{eqnarray}
Thus, in order to find the integrated heat kernel $K(t)$ one has to calculate the Laplace
transform of the function $\text{erfc}(u/(2\sqrt t))$. By making use of the table of the
Laplace transform~\cite{Erdelyi} one finds
\begin{equation}
\label{5-26} K(t)=\frac{1}{2}\,e^{tq^2}\,\text{erfc}\left ( q\sqrt t \right )
-\frac{1}{2}\,{.}
\end{equation}
The first term on the right-hand side of Eq.\ (\ref{5-26}) exactly reproduces Eq.\
(\ref{4-8}) while the constant term (-1/2) should be removed by introducing appropriate
renormalization when calculating the trace over the continuous variables.

In the TM-sector of the model under study the construction of the
local heat kernel proves to be much more complicated. First of
all, the factorization equation (\ref{5-2}) does not hold anymore.
However, the technique of integral equations~\cite{PNB} can be
applied here again, but the pertinent  integral equations will be
defined on the whole plane $z=0$, i.e., they are two-dimensional
integral equations. In order to go along this line, preliminary
one has to remove the dependence on the spectral parameter
$(\omega(p,\mathbf{k})/c)^2$ from the first matching condition
(\ref{2-12}). For the heat kernel $K(\mathbf{r},\mathbf{r};t)$
this condition can be transformed as follows
\begin{equation}
\label{5-27} \left [ \frac{\partial K}{\partial t}\,(\mathbf{s},z=0,\mathbf{r}';t)
\right]=2q\,\frac{\partial K}{\partial z}\,(\mathbf{s},z=0,\mathbf{r}';t)\,{,}
\end{equation}
where the notations $\mathbf{r}=(\mathbf{s},z),\;\mathbf{s}=(x,y)$, and (\ref{2-14}) are
used as before.

  The validity of this relation can be checked  easily if one takes into account that
the eigenfunctions $\varphi(\mathbf{r})\,e^{-\lambda_n t}$ entering the definition of the
heat kernel (\ref{5-1}) obey Eqs.\ (\ref{5-1b}) and (\ref{5-1c})
\begin{equation}
\label{5-29} -\Delta\, \varphi_n(\mathbf{r})\,e^{-\lambda_nt}=
\lambda_n\,\varphi_n(\mathbf{r})\,e^{-\lambda_nt}=-\frac{\partial}{\partial t}\,
\varphi_n(\mathbf{r})\,e^{-\lambda_nt}\,{.}
\end{equation}
Here the eigenvalues $\lambda_n$ are equal to $(\omega(p,\mathbf{k})/c)^2$. The matching
condition (\ref{5-27}) involves the space and time derivatives, i.e., so-called the skew
derivative.

The Green function $K(\mathbf{r},\mathbf{r}';t)$ obeys the heat conduction equations
(\ref{5-1b}) and (\ref{5-1c}), hence the matching condition (\ref{5-27}) can be obviously
rewritten in the form
\begin{equation}
\label{5-30} \left [ \Delta_{\mathbf{r}}K(\mathbf{s},z=0,\mathbf{r}';t) \right ]
=2q\,\frac{\partial K}{\partial z}\,(\mathbf{s},z=0,\mathbf{r}';t)\,{.}
\end{equation}

An interesting spectral problem arises here when we confine ourself to the
one-dimensional problem with $(\omega(p)/c)^2=p^2$: \begin{eqnarray}
-\frac{d^2}{dz^2}\,\varphi(z) &=&p^2\,\varphi(z)\,{,}\quad -\infty<z<\infty,\;z\neq0\,{,}
\label{5-30a}\\
\left [ \varphi'(z=0) \right] &=& 0\,{,}
\label{5-31}\\
\left [ \varphi''(z=0) \right] &=&2q\,\varphi'(z=0)\,{.} \label{5-32}\end{eqnarray} At
first sight, this spectral problem is completely different from the analogous one for the
$\delta$ potential (see Eqs.\ (\ref{2-11}), \ref{2-14a}), and \ref{2-13})). However the
respective eigenfunctions $\varphi_p(z)$ and $\Phi_p(z)$ are connected by relation
\begin{equation}
\label{5-33} \varphi'_p(z)=\Phi_p(z)\,{.}
\end{equation}
Both the problems have the same positive continuous spectrum
\begin{equation}
\label{5-34} 0<p^2<\infty\,{.}
\end{equation}
Furthermore the respective phase shifts, and consequently the scattering matrices,
coincide (see Eq.\ (\ref{2-20a}) and Eq.\ (\ref{2-21}) with $k=0$). From here we infer
immediately that these spectral problems have the same integrated heat kernels defined by
Eq.\ (\ref{4-8}), while the local heat kernels are obviously different.

The local heat kernel for the spectral problem (\ref{5-30a}) -- (\ref{5-32}) can be
derived in the same way as it has been done above in the spectral problem with $\delta$
potential by making use of the integral equations.

As far as we know, a couple of spectral problems possessing such
interesting features are found for the first time.
\section{Conclusion}
\label{sec6} The plasma sheet model investigated here proves to be  very interesting and
instructive, first of all, from the stand point of spectral analysis. The spectrum of the
model contains  both continuous branches and bound states (surface plasmon). It is
remarkable that for the latter the spectral zeta function cannot be constructed at least by
the standard analytic continuation method. At the same time the integrated heat kernel is
found  in an explicit form for all branches of spectrum. On the whole this heat kernel has
the asymptotic expansion of a noncanonical form  due to the singular point interaction.

By making use of the technique of integral equations developed by us earlier~\cite{PNB}
the local heat kernel in the TE-sector of the model is found in an explicit form. By the
way, here a new method is demonstrated for deriving the fundamental solution to the heat
conduction equation(or to the Schr\"odinger equation)  on an infinite line with the
$\delta $-like source. In principle the integral equations technique is also applicable
to construction of the local heat kernel in the TM-sector of the model.

For the heat equation on an infinite line with the $\delta$-source
a nontrivial counterpart is found, namely, a spectral problem with
point interaction, that possesses the same integrated heat kernel.
However the local heat kernels in these spectral problems are
different. The new boundary  problem is not selfadjoint, the
matching conditions at the singular point can be represent ed in
two forms: by equation containing the spectral parameter or ii) by
equation with second spatial  derivative (or the first time
derivative). As far as we know such a couple of spectral problems
is found at the first time.

Thus the spectral analysis of the model under consideration is  accomplished in full.

Without question, the  spectral analysis of the same model
describing the plasma layer of other forms, for example, circular
infinite cylinder or sphere, is also of interest.

 \acknowledgments
One of the authors, V.V.N., is indebted to Professor G.~Barton for valuable
communications concerning  the topics  considered in this paper.

 This research has been supported  in part by the
Heisenberg -- Landau Programme and by the Russian Foundation for
Basic Research (Grant No.\ 03-01-00025).

\end{document}